\documentclass[12pt]{article}
\usepackage{amsmath}
\usepackage{graphicx}
\usepackage{geometry}
\usepackage[applemac]{inputenc}

\usepackage[doublespacing]{setspace}

\ifx\pdfoutput\@undefined\usepackage[usenames,dvips]{color}
\else\usepackage[usenames,dvipsnames]{color}
\IfFileExists{pdfcolmk.sty}{\usepackage{pdfcolmk}}{} 
\fi
\usepackage[plainpages=false,pdfpagelabels,pagebackref=false,naturalnames=true,hyperindex=true,pdftitle={Living in Living Cities},pdfauthor={Carlos Gershenson}]{hyperref}
\hypersetup{colorlinks=true,
urlcolor=Cerulean,linkcolor=BrickRed,citecolor=RoyalBlue,
  pdfpagemode=None,
  pdfstartview=FitH}
\usepackage[all]{hypcap}

\begin{document}

\title{Living in Living Cities}
\author{Carlos Gershenson$^{1,2}$ \\
$^{1}$ Departamento de Ciencias de la Computaci\'on\\
Instituto de Investigaciones en Matem\'aticas Aplicadas y en Sistemas \\
Universidad Nacional Aut\'onoma de M\'exico\\
A.P. 20-726, 01000 M\'exico D.F. M\'exico\\
Tel. +52 55 56 22 36 19 \
Fax +52 55 56 22 36 20 \\
\href{mailto:cgg@unam.mx}{cgg@unam.mx} \
\url{http://turing.iimas.unam.mx/~cgg} \\
$^{2}$ Centro de Ciencias de la Complejidad \\
Universidad Nacional Aut\'onoma de M\'exico}
\maketitle

\begin{abstract}

This paper presents an overview of current and potential applications of living technology to some urban problems. Living technology can be described as technology that exhibits the core features of living systems. These features can be useful to solve dynamic problems. In particular, urban problems concerning mobility, logistics, telecommunications, governance, safety, sustainability, and society and culture are presented, while solutions involving living technology are reviewed. A methodology for developing living technology is mentioned, while supraoptimal public transportation systems are used as a case study to illustrate the benefits of urban living technology. Finally, the usefulness of describing cities as living systems is discussed.


\end{abstract}

\section{Urban Advantages and Disadvantages}

More than half of the world population lives in cities~\cite{Cohen:2003}. With 180,000 people moving to cities every day~\cite{Intuit:2010}, urban population is expected to grow, reaching $70\%$ of the global population by 2050~\cite{Cities:2010,Roberts:2011}. 

There are several advantages of urban settlements, such as less energetic requirements per capita, higher incomes, innovation, and productivity~\cite{Glaeser:2011,Bettencourt:2007,Bettencourt:2010}. In spite of---or perhaps because of---being highly attractive for people, modern cities also face several problems, such as mobility, crime, disease, pollution, and other social problems.

There have been several proposals concerning every urban problem, with different degrees of success. There are cities where the major problem might be mobility (Mexico City, Beijing~\cite{Gyimesi:2011}), safety (Ciudad Ju\'arez, Baghdad), unemployment (Detroit, Madrid), segregation (Chicago, Pretoria), traffic accidents (El Cairo, Dar-es-Salaam), or lack of infrastructure (Lagos, Kabul). Since there are different causes for different problems, there will be no single solution for all urban problems: several solutions have to be explored in parallel.

Urban planning has been guiding the development of cities for decades, at least in developed countries. Planning is certainly useful: it is better to deal with situations before they become problematic. However, urban planning has been rigid so far: how can future requirements be predicted as cities grow and embrace new technologies and customs? Just like a century ago cities were not planned to use cars as major means of transportation, cities cannot be planned now for their requirements of the next fifty years. Moreover, it is only recently that researchers have been able to develop descriptive models of urban growth~\cite{Andersson:2002b,Andersson:2002,Yamins:2003,Silva:2005,Mahiny:2012}.

The limitations of urban prediction are due to the complexity of cities~\cite{GershensonHeylighen2005,HeylighenEtAl2007,Gershenson:2011e}. Complexity implies that the components of a system are not separable. This lack of separability is due to relevant \emph{interactions} between components: The future state of components is co-determined by interactions, which cannot be enumerated, ordered, or predicted. Thus, prediction from initial and boundary conditions is limited. 

Cities can usefully be described as complex systems~\cite{portugali2000self,batty2005cities,Batty:2008}, since their components interact and co-determine their future. Thus, urban planning is limited~\cite{White2000High-resolution} by the very nature of their complexity. This does not imply that general properties of cities cannot be estimated, but that precise prediction is hopeless. To complement this lack of prediction, living technology can serve cities by providing a greater degree of adaptability and robustness.

In the next section, an overview of living technology and its properties is given. Section \ref{sec:urbanProblems} provides an extensive (although non-exhaustive) description of urban problems and solutions offered by living technology. In particular, problems in mobility, logistics, telecommunications, governance, safety, sustainability, and society and culture are discussed. Section \ref{sec:how} offers guidelines to develop urban living technology, using public transportation systems as a case study.
The paper concludes with a discussion on the usefulness of describing cities as living systems.

\section{Living Technology}

The term ``living technology" has been used to describe technology that is based on  the core features of living systems~\cite{Bedau:2009}. Living technology is adaptive, learning, evolving, robust, autonomous, self-repairing, and self-reproducing.  

\emph{Adaptation}~\cite{Holland1975,Holland1995} can be described as a useful change in a system in response to changes in its environment~\cite[p.19]{GershensonDCSOS}. Living systems are constantly adapting because their environment is dynamic. Adaptive technology is necessary where problems are dynamic. Certainly, there are different degrees of adaptation: a thermostat adapts only to changes of temperature, while an autonomous car has to adapt to changes in roads, traffic states, other vehicles, behavior of drivers, etc.

\emph{Learning} and \emph{evolution} can be seen as a second order adaptation, since they imply a permanent change in a system. In other words, after learning or evolution, a system will respond in a different way to similar circumstances.  Learning and evolution occur at different timescales: learning is a type of adaptation within a lifetime, while evolution is a type of adaptation across generations. Learning and evolving technologies are useful because they can adapt to novel circumstances. With these properties, the same system will be able to function in a broader range of situations. This increases the potential variety and complexity that the system can cope with~\cite{Ashby1956,BarYam2004,Gershenson:2010a}.

A system can be said to be \emph{robust} if it continues to function in the face of perturbations~\cite{Wagner2005}. Robustness---also called resilience---is prevalent in living systems and desired in technology~\cite{vonNeumann1956,Jen2005}, as it complements adaptation by allowing a system to ``survive" changes in the environment before it can adapt to them. Robustness and adaptation are deeply interrelated, since they are different ways to cope with unpredictable environments. Robustness is passive (changes are resisted by the system), while adaptation is active (changes cause a reaction in the system).  Robustness can be promoted by different properties~\cite{Gershenson:2010}, such as modularity~\cite{Simon1996,Watson2002,Schlosser:2004,Callebaut:2005,BalpoGershenson:2011}, degeneracy
\cite{Edelman:2001,FernandezSole2003,Wagner2004,Whitacre:2010}, and redundancy~\cite{GershensonEtAl2006}.

\emph{Autonomy}~\cite{Barandarian2004,MorenoRuiz2006,KrakauerZanotto2007} implies a certain independence of a system from its environment. Adaptation and robustness are requirements for autonomy, since they enable a system to withstand perturbations. Additionally, the autonomy of a system implies a certain degree of control over its own production~\cite{VarelaEtAl1974,McMullin2004} and behavior~\cite{Gershenson:2007}. Living systems have a high degree of autonomy. Technology has a tendency to become more and more autonomous of humans: from robots~\cite{Bekey:2005} to trading algorithms in stock markets~\cite{DeMarzo:2006}. This enables technology to respond to changes at faster rates. However, autonomous technology is also generating faster changes that affect other technologies.

\emph{Self-repair} and \emph{self-reproduction} can be seen as particular cases of \emph{self-organization}~\cite{GershensonHeylighen2003a}. Almost any system can be said to be self-organizing~\cite{Ashby1962}. However, it is \emph{useful} to describe a system as self-organizing when one is interested in relating how the interactions of elements affect the global properties of a system. This can be applied to living systems at several scales. For technology, self-organization can be used as an approach to build adaptive and robust systems~\cite{GershensonDCSOS}: interactions are designed so that elements find solutions by themselves. Thus, systems can adapt constantly to changes in their environment.

There cannot be a sharp distinction between non-living and living technology (just as there cannot be a sharp distinction between non-living and living systems). Nevertheless, it can be said that technology will be ``more living" as it has more and more of the core properties of living systems.

Living technology can be distinguished as primary or secondary~\cite[p. 91]{Bedau:2009}. \emph{Primary} living technology is constructed from non-living components, while \emph{secondary} living technology depends on living properties already present in its elements. Cities are secondary living technology, since living systems (humans, animals, plants, bacteria) are part of urban spaces. Nevertheless, the non-living components of cities have been acquiring with technology certain aspects of living systems, as mixed networks of soft, hard, and wet ALife~\cite[p. 92]{Bedau:2009}.

If cities always included living systems, have they always been using living technology? The answer depends on the deep question of the definition of life, which is far from being settled. To be able to decide ``how living" a system is, measures based on information theory can be used~\cite{Gershenson:2007}: we can measure how much the information of a system depends on the information of its environment. In this sense, ``more living" systems are those which are more autonomous from their environment, i.e. they produce more information about themselves than the information about themselves produced by their environment. Still, this measure depends on the scale at which the information is measured~\cite{GershensonFernandez:2012}. For example, it can be argued that a bacterium is more autonomous than a cell from a multicellular organism because it produces more of its own information. However, a multicellular organism produces more information about itself than a bacterial colony, since its organization at the multicellular scale can maintain its own integrity to a larger degree than the bacterial colony~\cite{Gershenson:2010a}. Thus, it can be argued that the organism is more autonomous than the colony at the multicellular scale.

If we are interested on deciding ``how living" urban technology is, we have to measure how much an urban system is able to produce its own information, which reflects its organization and thus control over its own dynamics, \emph{at the urban scale}. Different urban systems can be composed by the same living and non living components, e.g. traffic (drivers, pedestrians, vehicles, traffic lights, etc.). But different organizations of the urban system, e.g. traffic light coordination methods, will deliver different informational measures for the system, which will reflect their abilities to adapt, learn, evolve, and self-repair. For each urban system, if we increase its ``liveness" with living technology, the system will be able to deliver a better performance in comparison with a system without the properties of living systems.

\section{Solutions for Urban Problems}
\label{sec:urbanProblems}

Cities have been described metaphorically as organisms, e.g.~\cite{Dawson:1926,Spilhaus:1969}: they grow, have a metabolism, an internal organization, transportation networks of matter, energy, and information, and telecommunications have been characterized as ``nervous systems". Urban areas also reproduce and repair themselves, although their mechanisms are more akin to grasses than animals. Even thermodynamically, cities take matter, energy and information from their environment, transform them, and produce waste to maintain their organization, just like living systems.
 However, Lynch~\cite{Lynch:1981} argued that descriptions of modern cities as living organisms or as machines  are inadequate, even when they contain all twenty subsystems required by living systems, as defined by Miller~\cite{Miller1978Living-Systems}.
Still, the promise of living technology towards improving urban systems and thus transforming the nature of cities was not yet considered three decades ago.  Moreover, Batty~\cite{Batty:2012Cities} has recently argued that the scientific study of cities is transitioning ``from thinking of `cities as machines' to `cities as organisms'".

Bettencourt et al.~\cite{Bettencourt:2007} discovered that---in spite of several similarities---various properties of cities belong to different universality classes than those of biological organisms. Nevertheless, similar to living organisms, cities are constantly adapting~\cite{Bettencourt:2010}. In any case, this paper is not focussed on deciding whether cities are usefully described as living systems or not, but on exploring the use of living technology to solve urban problems.

Traditional approaches are efficient for \emph{stationary} problems, i.e. a solution is found, implemented, and the problem is solved. However, most 
urban problems are \emph{non-stationary}~\cite{Forrester1969Urban-Dynamics,Batty1971Modelling-Citie,Gershenson:2011b}: population changes with years, opinions can change within days, energy, resource, and waste requirements change with the seasons and with the hours of the day, traffic changes every second. Not only there are changes occurring constantly in urban spaces, but these occur at different scales. Solutions to these problems have to be robust and adapt, \emph{matching the scales} at which changes take place~\cite{GershensonDCSOS,Gershenson:2011a}.

Since urban problems are dynamic, urban technology has to find new solutions as problem changes by adapting, learning, and evolving. Living technology can offer this type of solutions~\cite{Mehaffy:2011,Alexander:2003}. Moreover, cities have been invaded by information technology~\cite{Kitchin:2011}, becoming a mesh of sensors, actuators, and controllers, exploiting the combined abilities of citizens and technology.

Biourbanism~\cite{williams1997biourbanism} has already proposed a similar path, looking at interdependencies between all the components of urban systems, focussing on sustainability and ecology. Biourbanism proposes the use of technologies that are closer to biology with the aim of having a reduced impact on the environment.

Information technology (IT) is bringing several properties of living systems to urban spaces~\cite{Kitchin:2011}. IBM's smart cities program aims at solving some urban problems with the aid of IT~\cite{Dodgson:2011,Harrison:2011}. The FuturICT european flagship project~\cite{Helbing:2011} proposes the integration of techniques from several disciplines to solve global problems, many of them urban. The Earth 2.0 project\footnote{\url{http://earth2hub.com}}  is also proposed at a global scale, using IT to build more adaptive and sustainable global and urban systems. The organic computing paradigm~\cite{OrganicComp:2011} focusses on information processing systems with properties of living systems. Organic systems can be considered as living technology.

In the next subsections, several urban problems and potential solutions with living technology are presented.

\subsection{Mobility}

The movement of people and goods is one of the major urban problems. It requires expensive infrastructure (roads, rails, ports, stations, bridges, vehicles, fuel, signalization). When mobility is inefficient or saturated, people lose time and money, gain stress, and more pollution is generated. Overall, the quality of life is reduced when mobility is limited or not efficient. There are several problems within urban mobility, so there will be no single solution for all of them~\cite{Cairns:2004}. At least eight interrelated aspects of urban mobility can be identified:

\begin{description}
\item[Transportation requirements. ] There is no mobility problem if people and goods do not have to be displaced. It is not possible for everyone to study, work and grow produce at home, but many actions can be taken to reduce the need of moving people and merchandise, i.e. the mobility demand.
\item[Scheduling. ] Congestion occurs when there are too many people in the same place at the same time. If people can transport themselves with more flexible schedules, then the demand of rush hours can dissipate over longer periods of time.
\item[Quantity. ] Too many vehicles or people saturate roads and public transportation systems. To reduce this, some cities use measures to demotivate use of private vehicles, such as high taxes, congestion charges, and limited parking. More flexible approaches to reduce vehicle quantity are carpooling and carsharing~\cite{Gansky:2010}, e.g. Zipcar and Buzzcar.
\item[Capacity. ] Building more and broader freeways, bike lanes, public transportation systems and efficient traffic lights increases the capacity of urban mobility. An increased capacity can be expensive, although technology can allow for increases in capacity at reduced costs.
\item[Behavior. ] Inadequate behavior of drivers or passengers can lead to delays in transportation. Examples for drivers include speeding, compulsive lane changing, and texting while driving. Examples for passengers include pushing and blocking, which can occur in different circumstances.
Potential interventions for restricting inadequate behaviors and promoting positive behaviors include education campaigns, fines, real-time information and social participation~\cite{singhal2008performance}.
\item[Infrastructure and technology. ] Infrastructure such as freeways, public transportation, bike lanes, and vehicle sharing systems can contribute to improve mobility. Technology can complement infrastructure, by enhancing its capacity. For example, traffic sensors can be used to coordinate traffic lights, avoid traffic jams, and suggest alternative routes.
\item[Society. ] In most societies, owning a car implies prestige, reflecting certain economic success. However, people are becoming so successful that roads are saturated. In several cities, people naturally prefer alternative modes of transport. With a social acceptance of  car-owning alternatives, it will become easier to balance different modes of transportation farther from private cars.
\item[Planning and regulation. ] Even when urban planning has limitations, cities suffer when there is no urban planning at all. In many cities this is complicated because politicians and not urbanists make the decisions on urban projects. Also, some cities do have planning and projects, but there is no enforcement nor regulation. Thus, plans never materialize and projects never are implemented. 
\end{description}

There are different actions that can be taken to improve different aspects of the eight factors mentioned above. For example, more capacity can be built. But if the quantity increases faster than the capacity, the improvement will be severely limited and problems will not be solved. In general, all of the eight factors have to be considered in parallel to improve urban mobility. In the next sections, examples of potential applications of living technology to address different problems in urban mobility are presented.

\subsubsection{Public Transportation}

When thousands or even millions of people have to move in urban areas through similar routes, mass transit becomes a better alternative than private motorized vehicles. Metro, bus rapid transit (BRT), trams, buses, and trains have been used since the nineteenth century for this purpose. 

According to theory, passengers arriving randomly at stations wait the least when headways---the temporal interval between vehicles---are equal~\cite{Welding:1957}. However, this configuration is always unstable, for all public transportation systems~\cite{GershensonPineda2009}. Random arrivals at stations will cause some stations to be busier than others. When a vehicle arrives at a busy station, it might be slightly delayed, increasing the headway with the vehicle ahead and reducing the headway with the vehicle behind. The longer headway might cause further delays at the next station, increasing even more the headway with the vehicle ahead and decreasing even more the headway with the vehicle behind. This equal headway instability leads to the formation of ``platoons" of vehicles that affect negatively the service, leading to long delays for passengers. There have been several approaches for dealing with equal headway instability in particular transportation systems~\cite{Turnquist:1980}.

Recently, it was found that transportation theory had misguided assumptions for decades~\cite{Gershenson:2011a}, namely that vehicles along a route will have the same travel time, thus developing methods that aim at maintaining equal headways, reducing waiting times for passengers \emph{at stations}. However, in order to maintain equal headways, some vehicles have to idle at stations. A self-organizing method was proposed~\cite{Gershenson:2011a}, where the equal headways are relaxed and even when passengers wait more at stations, the total travel times are reduced by a slower-is-faster effect~\cite{Helbing:2000,Helbing:2009}. The proposed method uses ``antipheromones" to make local decisions depending on neighboring vehicles and passenger demands at current stations, adapting to changing demands and delivering a ``supraoptimal" performance. The details of this solution are discussed as a case study in Section~\ref{sec:antiph}.

\subsubsection{Traffic Lights}

The coordination of traffic lights is an exponential-complete problem~\cite{PapadimitriouTsitsiklis1999,Lammer:2008}. Moreover, the traffic configuration changes constantly, as demands at intersections vary at the seconds scale. For this reason, fixed, optimizing approaches are limited for traffic light control~\cite{GershensonRosenblueth:2010}.

Adaptive methods, some of which are biologically-inspired, have been proposed to regulate traffic lights. Faieta and Huberman~\cite{FaietaHuberman1993} proposed an algorithm inspired in firefly synchronization, while Ohira~\cite{Ohira1997} proposed a controller based on an analogy with neural networks. 

Self-organizing traffic lights~\cite{Gershenson2005,HelbingEtAl2005,CoolsEtAl2007,Lammer:2008,Prothmann:2009,GershensonRosenblueth:2011,deGier:2011} can adapt to the local traffic demand, leading to an emergent and robust global coordination of traffic lights. Some of these methods are in the process of being implemented~\cite{Lammer:2010}, reporting considerable improvements in waiting times for cars, pedestrians, and public transport. This leads to economic, energetic, environmental, and social savings. 

\subsubsection{Real-time Information}
\label{sec:RTinfo}

The commercialization of GPS devices allowed drivers to query for the shortest route to their destination. However, once several people were using GPS, shortest routes were saturated, since everyone was advised to follow them. Shorter but not fastest. Real-time information---available for decades in radio traffic reports---can help drivers adapt their route according to the current traffic situation. A limitation of radio reports is that they are broadcasted: all drivers get the same information, most of which might not be relevant, and drivers cannot demand for particular information.
This scenario has changed in recent years, with applications such as Google Maps\footnote{\url{http://maps.google.com}} and Waze\footnote{\url{http://www.waze.com}}, which provide real time traffic information on demand. 

A key element of real-time information systems consists of sensors~\cite{Chong:2003,dressler2007selforg}. Once traffic states are detected, broadcasting or making them available is relatively straightforward. Since there are different types of sensors (fixed, mobile), 
sensor integration~\cite{Qi:2001} is a relevant problem to obtain useful information.

Intervehicle communication can provide useful real-time local information, which can be exploited to adapt to dynamic traffic states and improve traffic flow~\cite{Kesting:2008}.

Real-time information for public transportation systems can also help passengers to adapt their routes more efficiently, and even their behavior~\cite{GershensonPineda2009}.

In general, location-based services offer a broad application potential~\cite{Ratti2006Mobile-Landscap}.

\subsection{Logistics}
\label{sec:logistics}

Biologistics~\cite{Helbing:2009a} notes that the organization, coordination and optimization of various material flows is not restricted to artificial systems, but that living systems also have to deal with material flows. Moreover, living systems can handle material flows efficiently, adaptively, robustly, and learn from past experiences. Thus,  
with biological inspiration, using principles of modularity, self-assembly, self-organization, and decentralized coordination, artificial logistic systems can be designed that can adapt efficiently to changes of demand. 

A drawback to traditional approaches in logistics is that the supplies and demands for different goods are dynamic and unpredictable. This demands approaches where systems can adapt to changing demands at the same scale at which changes occur. 
For example, swarm intelligence~\cite{BonabeauEtAl1999,Kennedy:2001,Trianni-Tuci:09:ecal} has been applied to several problems in logistics~\cite{Svenson:2004}. 

Computationally, algorithms inspired by swarms or neurons are equivalent~\cite{Gershenson:2010b}, since they function at multiple scales, allowing them to compute solutions to problems at a faster scale and at the same time adapt to changes in problems at a slower scale. This is a desired property in logistics and many other areas.

\subsection{Telecommunications}   

A distinction can be made  between synchronous and asynchronous communicaiton~\cite{DesanctisMonge1999,GershensonSOBs}. IT has reduced delays of information transmission, allowing for technologies with faster response rates. Moreover,
IT has made it possible to shift from from broadcasted information to information on demand. Availability of information is a requirement for living urban technology, since relevant information is required in order to adapt, learn, and evolve. This was already illustrated in Section \ref{sec:RTinfo}.

Telecommunications have an essential role towards the use of living technologies in urban spaces. Not only for information transmission among citizens, but also among devices and systems~\cite{Resch:2011}. For this purpose, several approaches have been proposed to build adaptive, flexible and robust telecommunication networks~\cite{dressler2008bio-inspired,dressler2010bioinspired}. These networks are becoming so complex and operate at such speeds, that their technology can only function efficiently exhibiting the properties of living systems.

Telecommunication systems are relevant not only for transmission of information, but enable other uses of living technology in urban spaces, such as governance~\cite{pitroda1993development}.

\subsection{Governance}

Bureaucracies are often seen as rigid, slow, and inefficient. 
Living technology can enable the adaptive transmission of relevant information to govern cities~\cite{GershensonSOBs}. Any adaptive system requires sensors to be able to detect when changes are required. An obstacle in governance is that sensors are rather poor to allow governments to make informed decisions. Simply there is no infrastructure to detect what are the requirements of citizens. For example, India is connecting 250,000 local governments (Panchayats) to deliver and obtain information to and from citizens~\cite{Panchayats:2010}. 

Sensors are important but not the only aspect where changes are being made. Technology can also be used to make better collective decisions~\cite{RodriguezSteinbock2004,RodriguezEtAl2007}. This possibility enables societies to respond adaptively to different situations. This also helps governments to better administer cities. 

Governments have also been making their data publicly available, so that citizens can use this information in novel ways~\cite{Bizer:2009}. Opening data and information enables many potential applications. 
Also data created by citizens can be useful. For example, after the 2010 Haiti earthquake, people used OpenStreetMap\footnote{\url{http://www.openstreetmap.org/}} to improve maps and assist rescue and humanitarian aid efforts, identifying via satellite pictures collapsed buildings, refugee camps, and other damages.

The availability and processing of masses of urban data open the potential to governments that adapt constantly to changes in demand of their citizens. Moreover, they allow an increased citizen participation in governance, slowly fading the differences between governors and the governed. An extreme democracy can be reached when the opinion of every citizen has the same weight on any political affair. This would be achieved only with living technology, since such a system would have to adapt constantly to the changes in the population. 

\subsection{Safety}

In a similar way that living technology can improve governments, it can improve urban safety. On the one hand, prompt and adaptive response to natural and artificial catastrophes is facilitated. On the other hand, an urban mesh of sensors can increase public safety by monitoring public and private spaces, thus increasing citizen accountability. Simply having cameras to detect traffic infractions enforces people to comply with traffic rules, which---if designed properly---lead to increased road safety.

Artificial immune systems (AIS) have been proposed to to prevent intrusions in networked systems~\cite{hofmeyr1999immunity}. AIS exhibit properties of their biological counterparts: they are distributed, robust, dynamic, diverse and adaptive. Since intrusions are seldom repeated, security systems have to be flexible enough to adapt and respond to novel situations constantly. 

If used properly, living technology could also reduce crime rates. Having an effective police is not a solution for urban crime, since its causes seem to lie in unemployment, lack of opportunities, social influence, and several other factors~\cite{weatherburn2001causes}. Nevertheless, crime prevention is necessary, and it will be more effective if it exhibits properties of living systems~\cite{ekblom1999can}, since changing circumstances, trends and behaviors open constantly new niches for crime. Thus, an effective crime prevention has to adapt to these changes, to learn from previous experiences, and to be robust in the process. It might be just a coincidence, but life has become safer as technology has evolved~\cite{pinker2011better}. The causal relations between technology and safety have yet to be explored, but this trend probably will continue, increasing safety as technology becomes ``more living". 

\subsection{Sustainability}

Sustainability is the capacity to endure. For cities, sustainability involves not only environmental relations, but also economical and social. Material and energetic resources are required to ``fuel" cities, as well as economic and social benefits to attract and sustain citizens \cite{Trantopoulos2012Toward-Sustaina}.

Concerning material sustainability, pollution has to be considered. If there is less waste produced, then the complexity of waste management will be reduced. Cleaner and more efficient technologies can help in this direction. For example, if traffic flow is more efficient, less pollution will be produced by motor vehicles. Also, 
local production reduces transportation and transmission burdens, but the cost of production might be higher. Thus, a balance between mass production (cheaper to produce but distribution required) and local production (more costly to produce, cheaper to distribute) should be sought. Nevertheless, living technology can contribute to both reducing the cost of local production and to increase the efficiency of distribution (See Section \ref{sec:logistics}).

Synthetic biology~\cite{Benner:2005} (wet second-order living technology) is promising for producing cleaner fuels~\cite{Lee:2008}, as well as technology to reduce or prevent pollution, such as buildings that absorb carbon dioxide and bioluminescent trees that do not require electricity~\cite{Armstrong:2010}.

The efficient and adaptive production and distribution of energy, as envisioned by the concept of a ``smart grid"~\cite{gellings2009smart,Anghel:2007} is similar to other urban problems: there is a varying demand, as well as a varying production, which ideally should match the demand. Living technology can certainly benefit energy grids, coordinating local generation of energy and distributing it ``on demand".

Another application of living technology is the dynamic regulation of rainwater to collect water and prevent floods, where catchment systems react on the weather forecasts and water supply levels~\cite{Mims:2011,Ruhnke:2011}.

``Smart skins" for buildings have also been proposed for temperature regulation minimizing energy consumption.~\cite{Ritter:2007}.

A sustainable economy should produce more than what it consumes. Moreover, it has to accommodate employments, opportunities, and pensions for dynamic populations (aging in some countries, growing fast in others). W. Brian Arthur has recently described ``the second economy"~\cite{Arthur:2011}, based on information technology, where processes are interacting, adapting, and having an effect on the ``physical" economy. Arthur mentions that the second economy has properties of living systems, since digital devices and processes are starting to sense, compute, make decisions, and perform actions adaptively and independently of humans. 

Businesses and enterprises also have to develop and acquire living technology, since the demands of the markets are changing constantly and at increasing speeds. Organizations that are adaptive and robust will have better chances of enduring unpredictable changes in the economy. Moreover, urban living technology is itself a novel business niche~\cite{Arup:2011}.

Living technology can also have a positive effect in the social aspects of urban spaces. Safety was already mentioned, but in general living technology can help citizens to be more cooperative. Taking the example of driving, in some cases it might be beneficial for a driver to drive in such a way that affects negatively other drivers, tempting them to do the same. When a few drivers follow this behavior, the traffic becomes worse for everybody, including those that attempt to get a benefit. 
Cooperation has been extensively studied with game theory~\cite{Axelrod:1981,Nowak2006}. Living technology can provide several alternatives to promote cooperation. On the one hand, those who do not cooperate could be punished automatically. On the other hand, those who do cooperate could be rewarded. Moreover, living technology could help change situations in such a way that it will be beneficial for individuals to behave in such a way that is beneficial for the society as well. In other words, if the payoff for cooperating is always the highest, there will be no social dilemmas: everybody will selfishly cooperate.

\subsection{Society and Culture}

One example of a social benefit is given with innovation, which is already promoted by cities~\cite{Bettencourt:2007}. Can living technology accelerate innovation in cities? It seems that the answer is affirmative, at least indirectly: if living technology can solve at least some of the urban problems mentioned above, it will increase the attractiveness of cities to citizens. Moreover, it will increase the ``carrying capacity" of sustainable cities. Since larger cities tend to be more innovative, and living technology would allow cities to grow even more, it can be concluded that such ``living cities" will have an increased innovation rate. And innovation not only in science and technology, but in culture, education, and art as well.

Since IT and the Internet are reducing the burden of transportation, people are exchanging information remotely and globally, overflowing the benefits of urbanization across cities.

Social media---such as Twitter and Facebook---are transforming and facilitating social interactions. For example, ``Social moods" have been detected~\cite{Bollen:2011}. Technology over social networks could potentially be used to steer social behavior, for example preventing unhealthy habits and promoting healthy ones~\cite{Gershenson:2011c}.

\section{How to do it?}
\label{sec:how}
In the previous section, examples of existing and potential urban living technologies were mentioned. This section will focus on how living technology can be applied to urban problems.

Recently, a methodology was developed for designing and controlling systems that require to be adaptive and robust using the concept of self-organization~\cite{GershensonDCSOS}. Instead of designing a system to solve a problem that is changing constantly, with self-organizing systems  components are designed so that they find solutions by interacting among themselves. This allows them to \emph{autonomously} \emph{evolve}, \emph{learn} and \emph{adapt} to changes in the problem and to continue functioning in a \emph{robust} way. The methodology focuses on identifying the nature of interactions and eliminate or reduce negative interactions (``friction") and promote positive interactions (``synergy"~\cite{Haken1981}). Interaction improvement always leads to system performance improvement~\cite{GershensonDCSOS}. This approach is useful when the problem or situation is unknown, undefined, or dynamic.

The methodology is only one of several that have been proposed with similar aims in the literature. A review and comparison can be found in~\cite{Frei:2011}. Engineering methodologies that embrace complexity are promising for developing living technology. This is because they offer frameworks where artificial systems with the properties of living systems can be developed. 


In the next subsection, public transportation systems are used as a case study where living technology based on self-organization offers even better performance than the theoretical optimum.

\subsection{A case study: self-organizing public transportation systems}
\label{sec:antiph}

\begin{figure}[!ht]
\begin{center}
\includegraphics[width=4in]{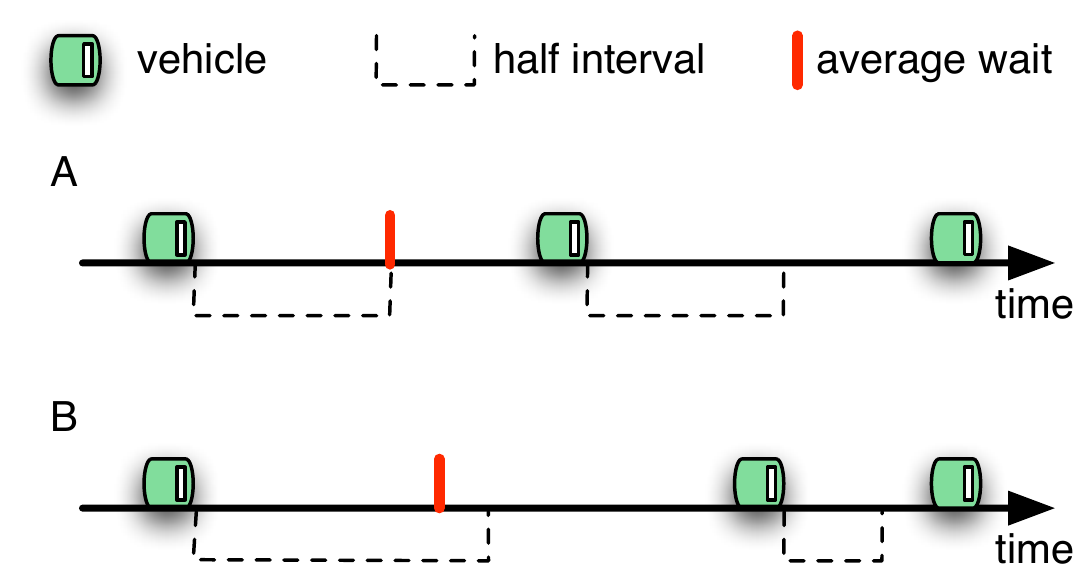}
\end{center}
\caption{
A. Equal headways lead to shorter passenger waiting times at stations. On average, the waiting time at stations is half the intervehicle interval. B. With unequal headways, passengers also are expected to wait half the current intervehicle interval, but there is a higher probability of passenger arrival within longer intervals, leading to higher average waiting times~\cite{Gershenson:2011a}.
}
\label{fig:headways}
\end{figure}

Passengers arriving randomly at stations will wait the least time if the headways (intervals between vehicles) are equal~\cite{Welding:1957}, as illustrated by Figure \ref{fig:headways}.

\begin{figure}[htbp]
\begin{center}
\includegraphics[width=15cm]{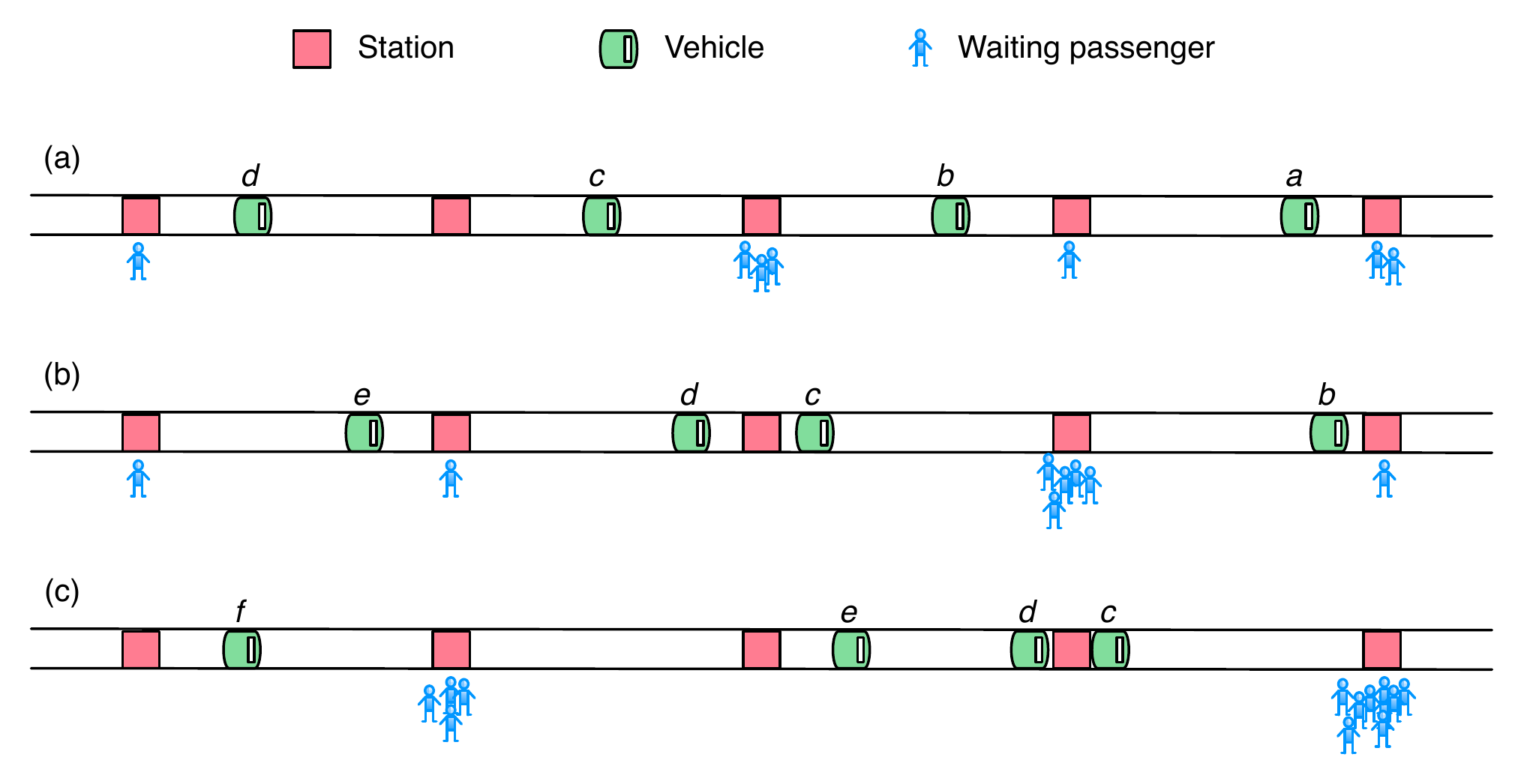}
\caption{Equal headway instability. a) Vehicles with a homogeneous temporal distribution, \emph{i.e.} equal headways. Passengers arriving at random cause some stations to have more demand than others. b) Vehicle $c$ is delayed after serving a busy station. This causes a longer waiting time at the next station, leading to a higher demand ahead of $c$. Also, vehicle $d$ faces less demand, approaching $c$. c) Vehicle $c$ is delayed even more and vehicles $d$ and $e$ aggregate behind it, forming a platoon. There is a separation between $e$ and $f$, making it likely that $f$ will encounter busy stations ahead of it. This configuration causes longer waiting times for passengers at stations, higher demands at each stop, and increased vehicle travel times. The average service frequency at stations is much slower for platoons than for vehicles with an equal headway~\cite{GershensonPineda2009}.}
\label{fig:HeadwayDeviation}
\end{center}
\end{figure}

Even when an equal headway configuration is desired, this is unstable, as explained in Figure \ref{fig:HeadwayDeviation}. It is like an inverted pendulum, where any perturbation kicks the system off balance and brings the pendulum down. In a similar way, public transportation systems ``prefer" to have unequal headways, as small delays amplify with a positive feedback, leading to the collapse of the system. Much of public transportation engineering for the past fifty years has dealt with trying to force transportation systems into maintaining an equal headway configuration~\cite{Turnquist:1980}.

In this context, a self-organizing method was developed with the aim of not only maintaining equal headways, but also to recover from unequal headway configurations. Following the inverted pendulum analogy, the goal was to build a system that would not only to prevent the pendulum from falling, but also to lift it up starting from a fallen position. 

Inspired by the adaptivity of ant communication~\cite{CamazineEtAl2003}, the method was tested and refined. One type of ant communication involves the secretion and sensing of pheromones. For example, if an ant finds a source of food, it will return to its nest with some food while leaving a pheromone trail. Other ants have a tendency to follow pheromone trails, proportional to the pheromone concentration. Thus, if more ants follow the pheromone trail and find the source of food, they will also return to the nest bringing food and reinforcing the pheromone trail, increasing the probability of recruiting more ants. Once the food source is exhausted, ants stop reinforcing the pheromone trail, which evaporates with time to prevent more ants from going to an empty source. Once a new source of food is found by exploring ants, new pheromone trails are formed. This communication via the environment is also known as ``stigmergy"~\cite{TheraulazBonabeau1999}. Functionally, the cognition of insect colonies mediated by stigmergy is analogous to neural cognition~\cite{Gershenson:2010b}.

The self-organizing method proposed for regulating public transportation headways is also stigmergic. However, instead of using pheromones, it uses ``antipheromones". Pheromones are placed by insects and evaporate with time, thus reducing their concentration. Antipheromones are virtual markers that increase their concentration with time, while they are erased by passing vehicles. A simple algorithm determines how much time each vehicle should spend at each station depending on the amount of \emph{passengers} waiting at the station, the antipheromone concentration (which is directly proportional to the \emph{time} since the  vehicle ahead departed), and the \emph{distance} to the vehicle behind~\cite{Gershenson:2011a}. This algorithm enables each vehicle to adapt to the demand at each station, preventing idling that occurs when equal headways are maintained, and allows enough robustness to prevent the platooning of vehicles and flexibility to recover from platooned configurations.

Discrete computer simulations were performed to compare the self-organizing method with a ``default" method, which does not restrict any waiting time and always leads to equal headway instability, and an ``adaptive maximum" method~\cite{GershensonPineda2009}, where there is a minimum waiting time at stations for vehicles and a maximum waiting time is modified depending on the global passenger demand and headways are always maintained, but not recovered. In the simulations, vehicles have a maximum passenger capacity and move discretely one space unit per time step, unless there is another vehicle ahead, passengers are boarding or descending at stations, or there is another restriction, such as waiting times at stations. Passengers arrive randomly at stations with a Poisson distribution on average every $\lambda$ time steps. When a vehicle arrives at a station, passengers scheduled to descend exit taking one time step each. Then passengers waiting at the station board taking one time step each until the vehicle is full or leaves the station. 

\begin{figure}[!ht]
\begin{center}
\includegraphics[width=3.5in]{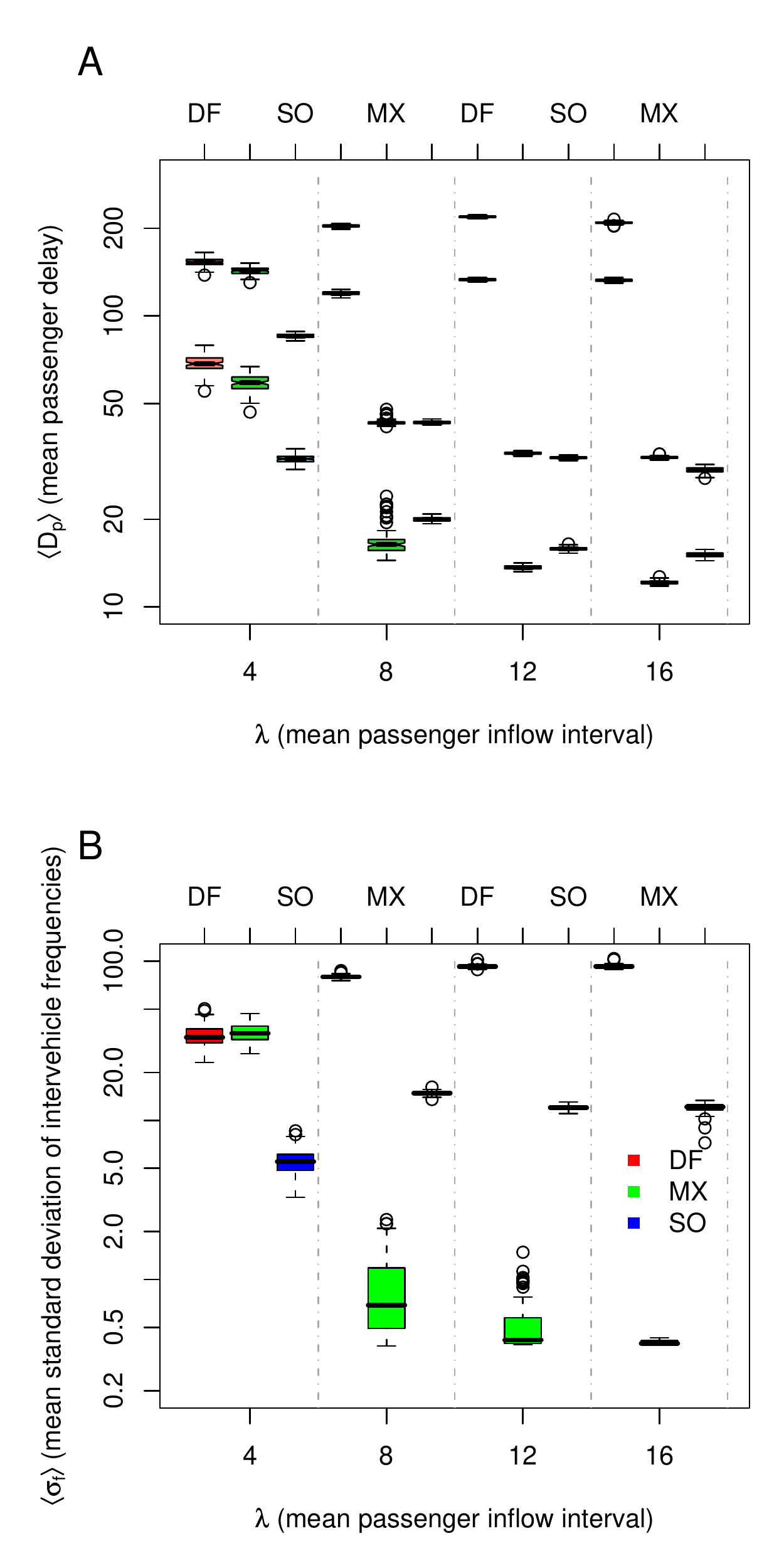}
\end{center}
\caption{
Results for homogeneous scenario.  A. Passenger delays for methods: ``default" (\emph{DF}), ``max" (\emph{MX}), and ``self-organizing" (\emph{SO}), for different passenger demands (lower $\lambda$ means higher demand). Lower boxes at each column show waiting times at stations. Higher boxes show total waiting times. B. Headway standard deviations. Lower $\sigma_f$ implies more regular headways. \emph{DF} shows unstable headways, \emph{MX} equal headways (except for $\lambda=4$), and \emph{SO} adaptive headways. Notice logarithmic scale~\cite{Gershenson:2011a}. 
}
\label{fig:homo}
\end{figure}

Results for a homogeneous scenario, with equidistant stations and initial positions of vehicles and equal passenger demand ($\lambda$) at stations, are shown in Figure \ref{fig:homo} for four different passenger demands. The headways in the default method collapse (as seen by the high standard deviations of intervehicle frequencies) leading to very high waiting times.
Surprisingly, the self-organizing method, even when headways are not maintained (although the system does not collapse), produced waiting times even lower than those of the maximum method, which maintained equal headways. Theory would tell us that waiting times are optimal for an equal headway configuration, meaning that the self-organizing method delivers supraoptimal performance. Still, when passenger waiting times are separated between total waiting times and waiting times at stations, the maximum method indeed has the minimum waiting times at stations, which is what the theory tells us. However, the theory assumes that travel times are independent of waiting times at stations, and they are not. In order to keep equal headways, some vehicles must idle, while others must leave some passengers behind. The self-organizing method is flexible enough so that headways are not maintained but neither collapsed, while passengers at stations are served on demand. Thus, even when waiting times at stations are higher, the total waiting times are lower.

\begin{figure}[!ht]
\begin{center}
\includegraphics[width=3.5in]{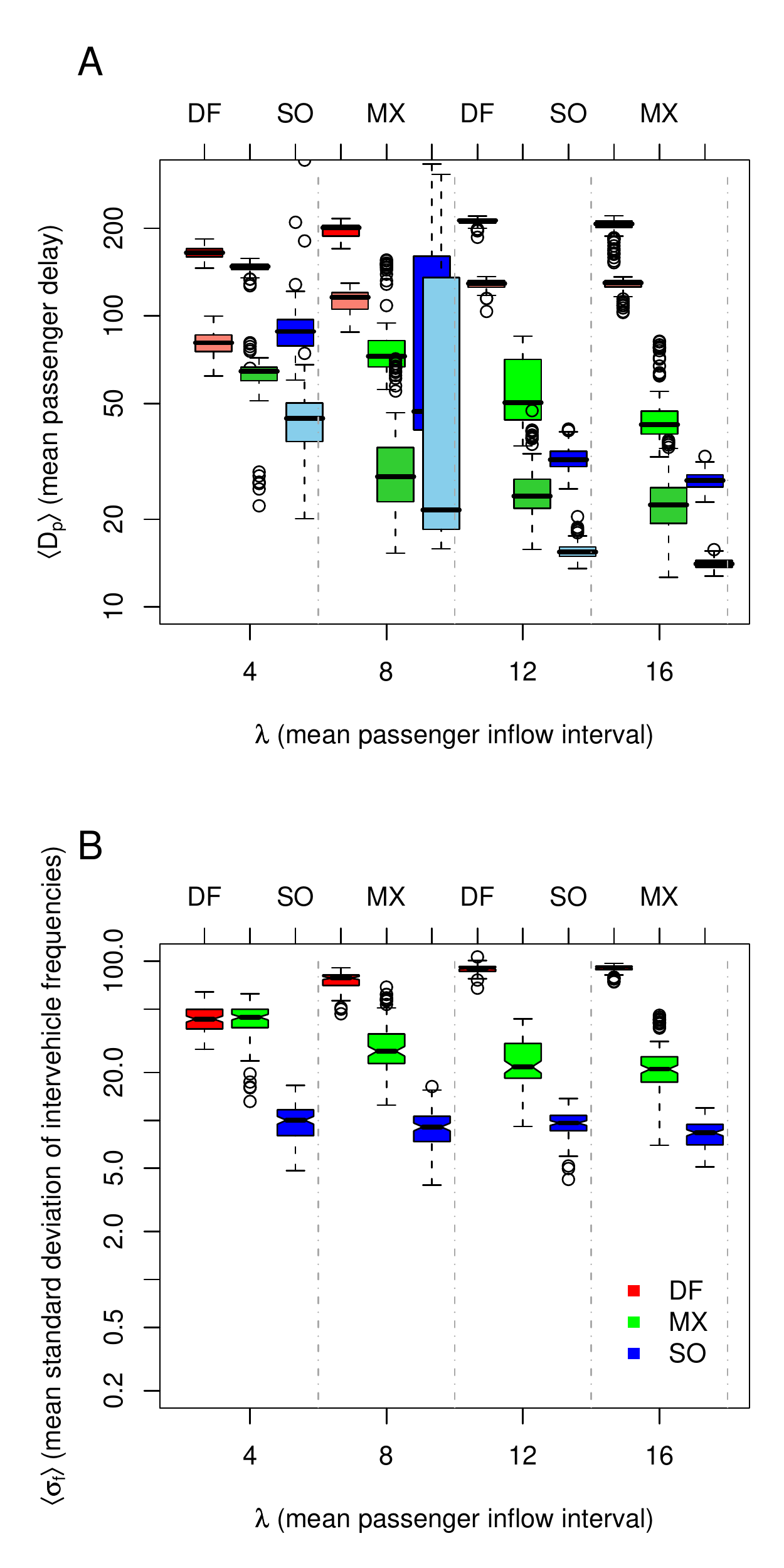}
\end{center}
\caption{
Results for non-homogeneous scenario. A. Passenger delays for methods: ``default" (\emph{DF}), ``max" (\emph{MX}), and ``self-organizing" (\emph{SO}), for different passenger inflow intervals $\lambda$. Lower boxes, slightly shifted to the right, at each column show waiting times at stations. Higher boxes show total waiting times. B. Headway standard deviations. Lower $\sigma_f$ implies more regular headways. Notice logarithmic scale~\cite{Gershenson:2011a}.
}
\label{fig:nonhomo}
\end{figure}

Results for a non-homogeneous scenario, with non equidistant stations nor initial positions of vehicles and unequal passenger demand ($\lambda$) at stations, are shown in Figure \ref{fig:nonhomo}. The default method collapses as well. The maximum method is not able to recover from the unequal initial headways and maintains them, leading also to high waiting times, even at stations, although not as high as for the default method. The self-organizing method is able to adapt to the non-homogeneous demands in this scenario and delivers a performance similar to that of the homogeneous scenario.

The self-organizing method is better than the theoretical optimum because of a slower-is-faster effect~\cite{Helbing:2000,Helbing:2009}. Passengers indeed wait more time at stations, but trying only to minimize passenger waiting time by forcing equal headways leads to \emph{friction} between vehicles, since vehicles serving stations with different passenger demands will idle and or leave passengers unattended at stations. Passenger inflow is not predictable, and assuming average flows to force predefined schedules will also lead to friction because of the same reason. On the contrary, the self-organizing method promotes \emph{synergy} by stigmergy of the vehicles, since they can balance---communicating through the antipheromones---the load of the system without idling and without collapsing, adapting to the current passenger demand at every station and state of the vehicles. These positive interactions allow the reduction of travel times, which benefit vehicles and passengers.

Traditional public transport regulation is more like clockwork, attempting to impose equal headways to changing demands. The self-organizing method is more like a healthy heart, where different intervals adapt to the instant demands of the system. Our current public transportation systems are more like diseased hearts: either too regular (cannot adapt) or arrhythmic (inefficient).

As this case study showed, living technology (adaptive, robust, self-organizing) can deliver a higher efficiency than that of traditional systems. Solutions to urban problems require the properties of living systems because problems are constantly changing. This limits their predictability and thus solutions which are unable to adapt to unforeseen situations. Since living systems make a living out of adapting to unforeseen situations, living technology is an excellent candidate for solving urban problems.

\section{Beyond the Metaphor: Towards Living Cities}

Cities will offer a higher quality of life if they exhibit the properties of livings systems.
After listing several current and potential urban living technologies, one can ask to what extent speaking about living cities is a mere metaphor and to what extent cities are usefully described as living systems. 

Living systems are constantly adapting, learning and evolving because their environment is always changing at different timescales. Living systems also require to be robust to endure unforeseen perturbations. Efficient cities have to do the same. It is not enough being ``smart". The demands and conditions of cities change constantly at different scales, so they must adapt, learn and evolve in a robust fashion in order to endure. Cities are not physically similar to living systems (no DNA, no membranes), but functionally, they should exhibit the same properties. From a materialist point of view, it makes no sense to speak about living cities. However, from a functionalist point of view, it is very relevant to speak about the relationships between living systems, artificial life, living technology, and urban systems. This is because the properties of living systems (natural or artificial) can be exploited to solve urban problems, making cities more adaptive and robust. 

If a notion of life based on entropy or information is used~\cite{Adami:1998,Gershenson:2007}, then one can even \emph{measure} to what extent different cities can be considered to be alive, with a continuous transition between non-living and living systems~\cite{Bedau1998}. In non-technical terms, if a city has a sufficient control over its own production, endowing it with a certain autonomy and integrity, then it can be usefully described as a living system. Living technology has been contributing to the increase of the ``liveness" of cities, as it was shown by the examples presented in this paper. Moreover, the study of ``living cities" is related to at least one of the open problems in artificial life~\cite{BedauEtAl2000}: To determine whether fundamentally novel living organizations can exist.

Technology has always evolved~\cite{Kelly:2010}, but with the aid of humans for most of its history. As living technology is developed, technology will be able not only to be more adaptive and robust, but to evolve by itself in directions that we cannot foresee. What can be said is that the integration between technology and living systems---including humans---will increase. Living cities will be the outcome of this integration.
	
Will solutions to urban problems using living technology bring new problems? Since predictability is limited, most probably new problems will arise. Nevertheless, we can always transform problems into opportunities. How? By deciding to do something about them.

\section*{Acknowledgments}

I should like to thank Steen Rasmussen and two anonymous referees for useful comments. 
This work was partially supported by UNAM-DGAPA-IACOD project T2100111,  by Intel\textsuperscript{\textregistered}, and SNI membership 47907 of CONACyT, Mexico. 

\bibliographystyle{alj}
\bibliography{carlos,traffic,sos,complex,RBN,information,computing,swarmCognition,orgs,evolution}

\end{document}